\documentclass[prd,preprint,showpacs,preprintnumbers,nofootinbib,eqsecnum,superscriptaddress]{revtex4}

 \usepackage[dvips,final]{graphicx}
  \usepackage{amssymb}
   \usepackage{amsmath}
    \usepackage{amsfonts}
     \usepackage{epsfig}
      \usepackage{bm}

\usepackage{mathpazo}

\usepackage[section]{placeins}

\usepackage{multirow}
\usepackage{ctable}
\usepackage{booktabs}
\usepackage{array}
\usepackage{tabularx}
\usepackage{xcolor}
\usepackage{pstricks}

\begin{document}

\title{Searching for and exploring double-parton scattering effects \\in four-jet production at the LHC}

\author{Rafa{\l} Maciu{\l}a}
\email{rafal.maciula@ifj.edu.pl} 
\affiliation{Institute of Nuclear Physics PAN, PL-31-342 Cracow, Poland}

\author{Antoni Szczurek\footnote{also at University of Rzesz\'ow, PL-35-959 Rzesz\'ow, Poland}}
\email{antoni.szczurek@ifj.edu.pl}
\affiliation{Institute of Nuclear Physics PAN, PL-31-342 Cracow, Poland}

\date{\today}

\begin{abstract}
We discuss four-jet production at the LHC.
We calculate cross section for both single-parton scattering (SPS)
using the ALPGEN code as well as for double-parton scattering (DPS)
in leading-order collinear approach. Our results are compared
with experimental data obtained recently by the CMS collaboration.
We show that the ALPGEN code relatively well describes
distributions in transverse momenta and rapidity of each of the four
jets ordered by their transverse momenta (leading, subleading etc.).
The SPS mechanism does not explain the distributions at
large rapidity for the leading jet. The DPS mechanism considerably
improves the agreement with the experimental data in this corner of 
the phase space.
In order to enhance the relative DPS contribution
we propose to impose different cuts.
The relative DPS contribution increases when decreasing the lower
cut on the jet transverse momenta as well as when a low lower cut
on the rapidity distance between the most remote jets is imposed.
We predict very flat distribution in azimuthal angle between the
most remote jets with low lower cuts on jets transverse momentum.
We identify phase-space corners where the DPS content is enhanced
relatively to the SPS one.
\end{abstract}


\maketitle

\section{Introduction}

The physics of multiparton interactions (MPI), in general, or 
double parton scattering (DPS) for hard processes gained recently 
a new impulse related to the experiments at Large Hadron Collider (LHC) 
\cite{Bartalini2011,DPS_CERN,DPS_Antwerpen,MPI2014}.

The DPS contributions were discused for several reactions
to mention double Drell-Yan \cite{Halzen87,Kom2011}, creation of 
$b \bar b b \bar b$ \cite{Treleani_bbbarbbbar}, etc.
It was predicted \cite{LMS2012} and recently found \cite{LHCb_D0D0,MS2013} 
that the double parton scattering leads to very abundant production
of (at least) two $c \bar c$ pairs at the LHC.
Another process where DPS seems to be seen is double $J/\Psi$
production in the region of large rapidity separation between
the two $J/\psi$'s \cite{Stirling,Baranov2013}.

We will not discuss in this Letter evident progress 
in theoretical understanding of the DPS or MPI interaction in general, 
which took place in recent years. In contrast to very quantitative
and precise studies of many SPS perturbative partonic processes 
the field of MPI and DPS lacks similar precision.
So far most of the experimental studies have concentrated on the extraction
of $\sigma_{eff}$ in so-called factorized ansatz\footnote{In some cases
  even experimental extraction of $\sigma_{eff}$ may be subtle \cite{SeymourLHCb}.}. 
In our opinion, at present some efforts should be made in order to make 
the studies of DPS processes really quantitative and differential in
kinematical variables.
In order to make more detailed phenomenological studies one needs 
also more clear cases where the DPS effects could be identified without
any doubts.

Four-jet production was traditionally discussed in the context of
double parton scattering. Actually it was a first process where
the DPS was claimed to be observed experimentally \cite{Tevatron_4jets}.
However, in most of the past as well as current analyses the DPS
contribution to four-jet production is relatively small and single
parton scattering (SPS) driven by the 2 $\to$ 4 partonic processes
dominates. To pin down the DPS contribution one has to make rather
complicated studies comparing Monte Carlo templates and experimental
data. In such a case the result is not very transparent and
one may worry whether the final result is not an artifact of an inadequate
Monte Carlo generator. A more evident result for DPS in four-jet
production  would clearly provide a new impulse for the MPI community.

On the theoretical side the DPS effects in four-jet production were 
discussed in Refs.~\cite{Treleani_jet1,Treleani_jet2,Mangano,Domdey,Berger,Strikman}.
A first theoretical estimate of SPS four-jet production, including only
some partonic subprocesses, and its comparison to DPS contribution was
presented in Ref.~\cite{Mangano} for Tevatron.
Some new kinematical variables useful for identification of DPS
were proposed in Ref.~\cite{Berger}.
A model dependence on collision energy and minimal transverse energy
of $\sigma_{eff}$ was studied in Ref.~\cite{Domdey}. Presence of perturbative
parton splitting mechanism in the context of four-jet production
was discussed in Ref.~\cite{Strikman}.

In our recent studies we have shown how big can be the contribution of
DPS for (two) jets widely separated in rapidity \cite{Maciula:2014pla}.
Understanding of this contribution is important in the context of 
searching for BFKL effects or in general QCD higher-order effects
\cite{Chatrchyan:2013qza}. We have found that with the present cuts used 
in the CMS analysis \cite{CMS:2013eda} the DPS contribution can be 
of the order of 10-20\%.
It could be still somewhat enhanced imposing further
cuts on transverse momentum of the dijets or azimuthal angle between
jets.

In the present letter we wish to explore exclusive four-jet sample
where the situation in the context of searching for DPS should be 
even better. We shall discuss how to maximize the DPS contribution
by selecting relevant kinematical cuts.

\section{Theoretical formalism}

Partonic cross sections used to calculate DPS are only in leading order.
The cross section for dijet production can be then written as:
\begin{equation}
\frac{d \sigma(i j \to k l)}{d y_1 d y_2 d^2p_t} 
= \frac{1}{16 \pi^2 {\hat s}^2}
\sum_{i,j} x_1 f_i(x_1,\mu^2) \; x_2 f_j(x_2,\mu^2) \;
\overline{|\mathcal{M}_{i j \to k l}|^2} \;.
\label{LO_SPS}
\end{equation}
In our calculations we include all leading-order $i j \to k l$ partonic 
subprocesses.
The $K$-factor for dijet production is rather small, of the order of 
$1.1 - 1.3$ (see e.g. \cite{K-factor1,K-factor2}), 
but can be easily incorporated in our calculations. It was shown that
already the leading-order approach gives results in sufficiently reasonable 
agreement with recent ATLAS and CMS inclusive jet data \cite{Maciula:2014pla}.

This simplified leading-order approach can be however used easily in 
calculating DPS differential cross sections. 
The multi-dimensional differential cross section can be written as:
\begin{equation}
\frac{d \sigma^{DPS}(p p \to \textrm{4jets} \; X)}{d y_1 d y_2 d^2p_{1t}
  d y_3 d y_4 d^2p_{2t}} 
= \sum_{i_1,j_1,k_1,l_1;i_2,j_2,k_2,l_2} \; 
\frac{\mathcal{C}}{\sigma_{eff}} \;
\frac{d \sigma(i_1 j_1 \to k_1 l_1)}{d y_1 d y_2 d^2p_{1t}} \; 
\frac{d \sigma(i_2 j_2 \to k_2 l_2)}{d y_3 d y_4 d^2p_{2t}}\;, 
\label{DPS}
\end{equation}
where
$\mathcal{C} = \left\{ \begin{array}{ll}
\frac{1}{2}\;\; & \textrm{if} \;\;i_1 j_1 = i_2 j_2 \wedge k_1 l_1 = k_2 l_2\\
1\;\;           & \textrm{if} \;\;i_1 j_1 \neq i_2 j_2 \vee k_1 l_1 \neq k_2 l_2
\end{array} \right\} $ and partons 
$j,k,l,m = g, u, d, s, \bar u, \bar d, \bar s$. 
The combinatorial factors include identity of the two subprocesses.
Each step of DPS is calculated in the leading-order approach 
(see Eq.(\ref{LO_SPS})).

Experimental data from Tevatron \cite{Tevatron1,Tevatron2} and 
LHC \cite{Aaij:2011yc,Aaij:2012dz,Aad:2013bjm} 
provide an estimate of $\sigma_{eff}$ in the denominator of formula 
(\ref{DPS}). In the calculations we have taken in most cases 
$\sigma_{eff}$ = 15 mb.
Phenomenological studies of $\sigma_{eff}$ are summarized e.g. in
\cite{Seymour} with the average value $\sigma_{eff} \approx$ 15 mb.

Now we proceed to the SPS production mechanisms of four-jet production.
The elementary cross section for the SPS mechanism
has the following generic form:
\begin{equation}
d\hat{\sigma}_{ij \to klmn} = \frac{1}{2\hat{s}} \; 
\overline{|{\cal M}_{i j \rightarrow k l m n}|^2} \; d^{4} PS ,
\label{elementary_cs}
\end{equation}
where the invariant phase space reads:
\begin{equation}
d^{4} PS = \frac{d^3 p_1}{2 E_1 (2 \pi)^3} \frac{d^3 p_2}{2 E_2 (2 \pi)^3}
           \frac{d^3 p_3}{2 E_3 (2 \pi)^3} \frac{d^3 p_4}{2 E_4 (2 \pi)^3}
           (2 \pi)^4 \delta^4 \left( p_1 + p_2 + p_3 + p_4 \right) \; . 
\label{four_body_phase_space}
\end{equation}
Above $p_1, p_2, p_3, p_4$ are four-momenta of outgoing partons (jets).
Many possible subprocesses $i j \to k l m n$ are possible in general.
In some corners of the phase space only some processes are really
important and others may be safely neglected.

The hadronic cross section is then given by the integral
\begin{eqnarray}
d \sigma &=& \int d x_1 d x_2 \sum_{ijklmn}
           f_i(x_1,\mu_F^2) f_j(x_2,\mu_F^2) 
           \; d \sigma_{ij \to k l m n} 
      \; .
\label{hadronic cross section}
\end{eqnarray}
Above $f_i$ and $f_j$ are parton (gluon, quark, antiquark)
distribution function. 

Instead of explicitly using the formulae above we shall use
its Monte Carlo version as implemented in the event generator ALPGEN
\cite{ALPGEN}. Weighted events from the generator will be used to
construct distributions presented in the next section. 
Only light quarks/antiquarks are included in our
calculation.

\section{First results}

\begin{figure}[!h]
\begin{minipage}{0.47\textwidth}
 \centerline{\includegraphics[width=1.0\textwidth]{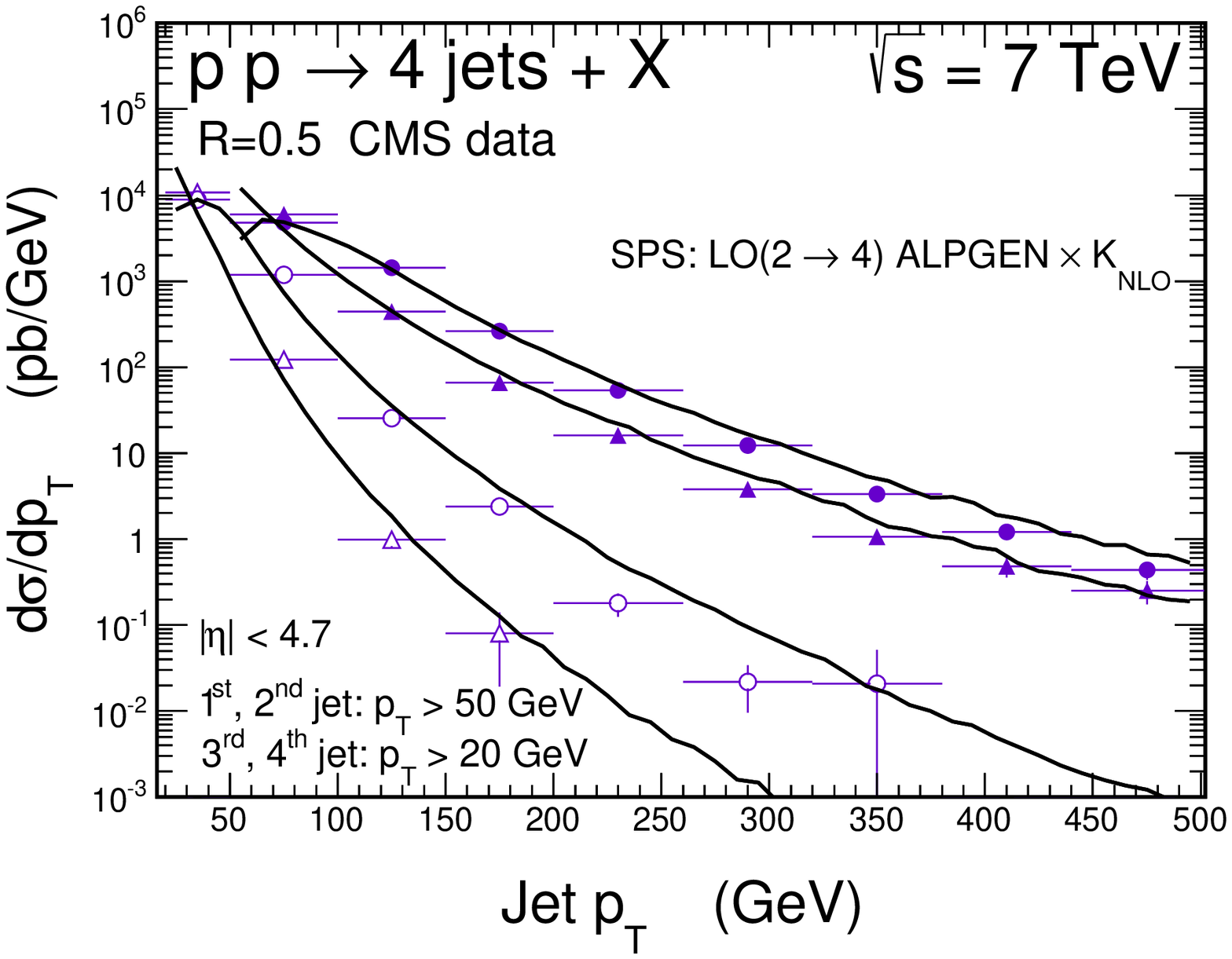}}
\end{minipage}
\hspace{0.5cm}
\begin{minipage}{0.47\textwidth}
 \centerline{\includegraphics[width=1.0\textwidth]{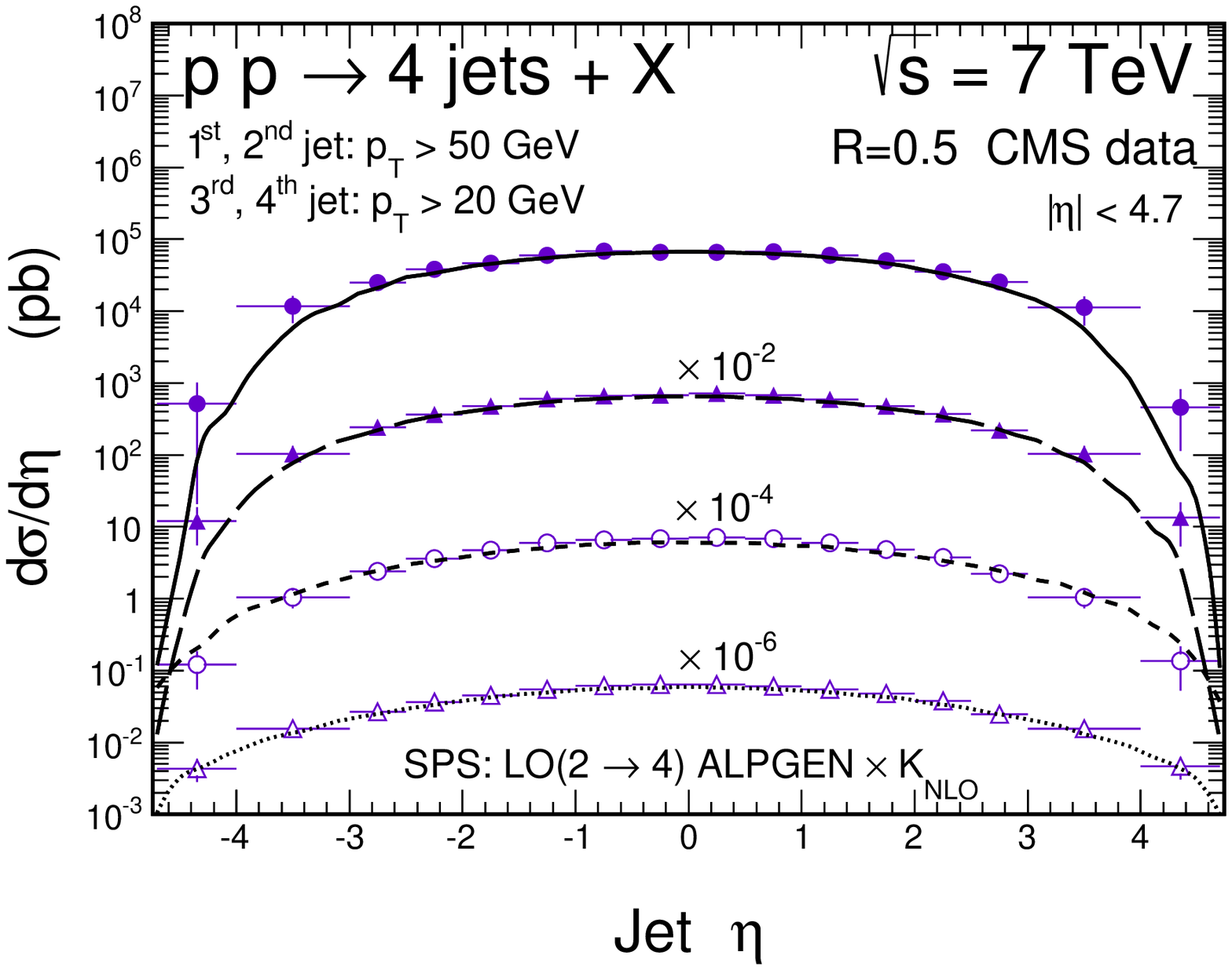}}
\end{minipage}
   \caption{
\small Transverse momentum (left panel) and rapidity (right panel) 
distributions of each of the four-jets (ordered in transverse momentum) 
in the four-jet sample together with the CMS experimental data 
\cite{Chatrchyan:2013qza}. The calculations were performed with
the ALPGEN code \cite{ALPGEN}. Here kinematical cuts relevant for
the experiment were applied to allow for a comparison.}
 \label{fig:CMS-data-1}
\end{figure}

To start our analysis we wish to check how reliable our SPS 
four jet calculation is. In Fig.~\ref{fig:CMS-data-1} we confront
the results of calculation with the leading-order code ALPGEN \cite{ALPGEN} 
with recent CMS experimental data \cite{Chatrchyan:2013qza}.
In this analysis the CMS collaboration imposed different transverse momentum
cuts on the leading, subleading, 3$^{rd}$ and 4$^{th}$ jets 
(see the figure caption). 
In this calculation we have used an extra $K$-factor to effectively 
include higher-order effects. How big is the $K$-factor was discussed
recently in \cite{Bern:2011ep}.
It was found that the next-to-leading order contributions are
important and lead to $K <$ 1. In our calculation here we will use
the $K$-factor found there.
We get relatively good description of both
transverse moentum and pseudorapidity distributions
of each of the four (ordered in transverse momentum) jets.
Therefore we conclude that the calculation with the ALPGEN generator 
can be a reliable SPS reference point for the DPS effects.

\begin{figure}[!h]
\begin{minipage}{0.47\textwidth}
 \centerline{\includegraphics[width=1.0\textwidth]{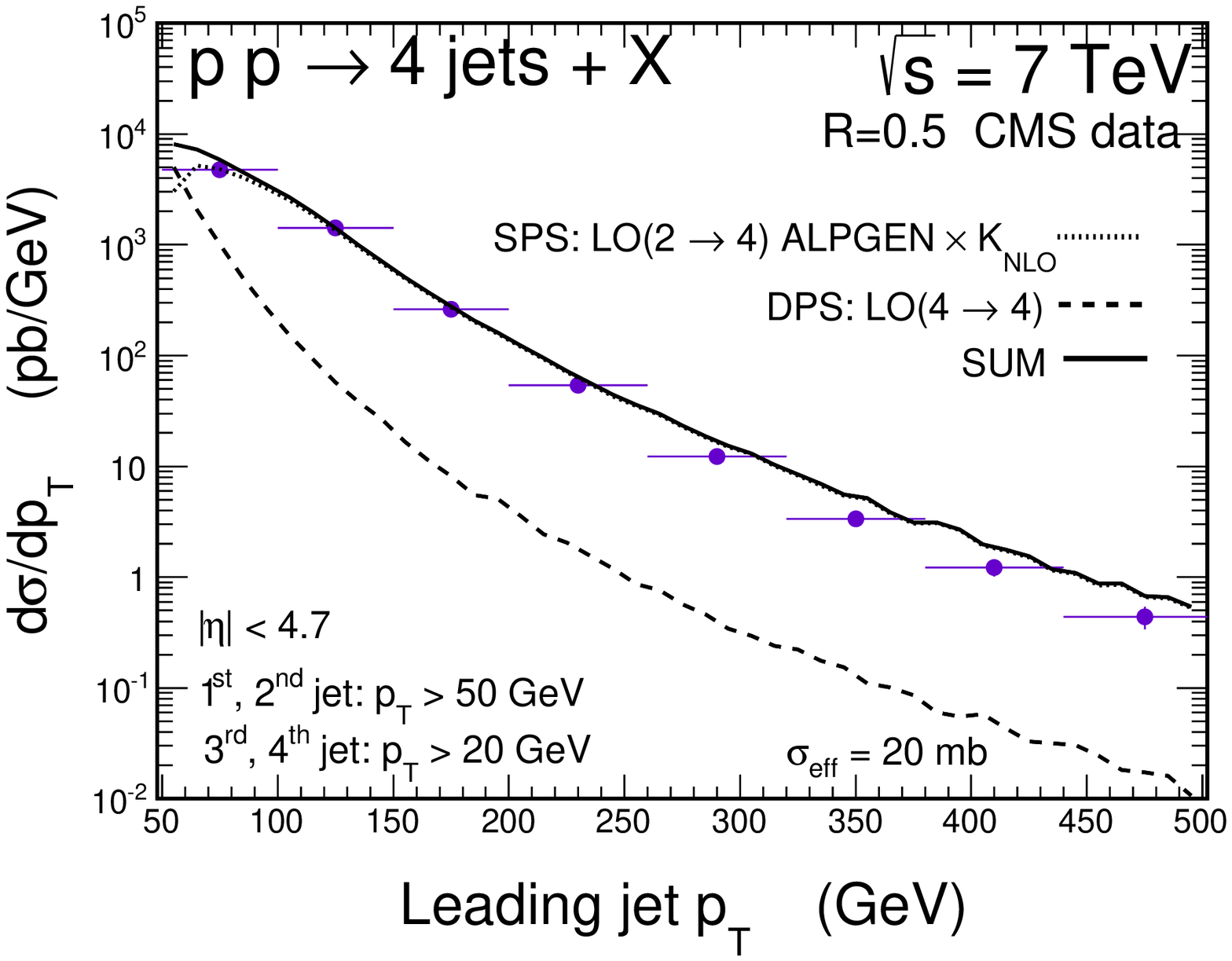}}
\end{minipage}
\hspace{0.5cm}
\begin{minipage}{0.47\textwidth}
 \centerline{\includegraphics[width=1.0\textwidth]{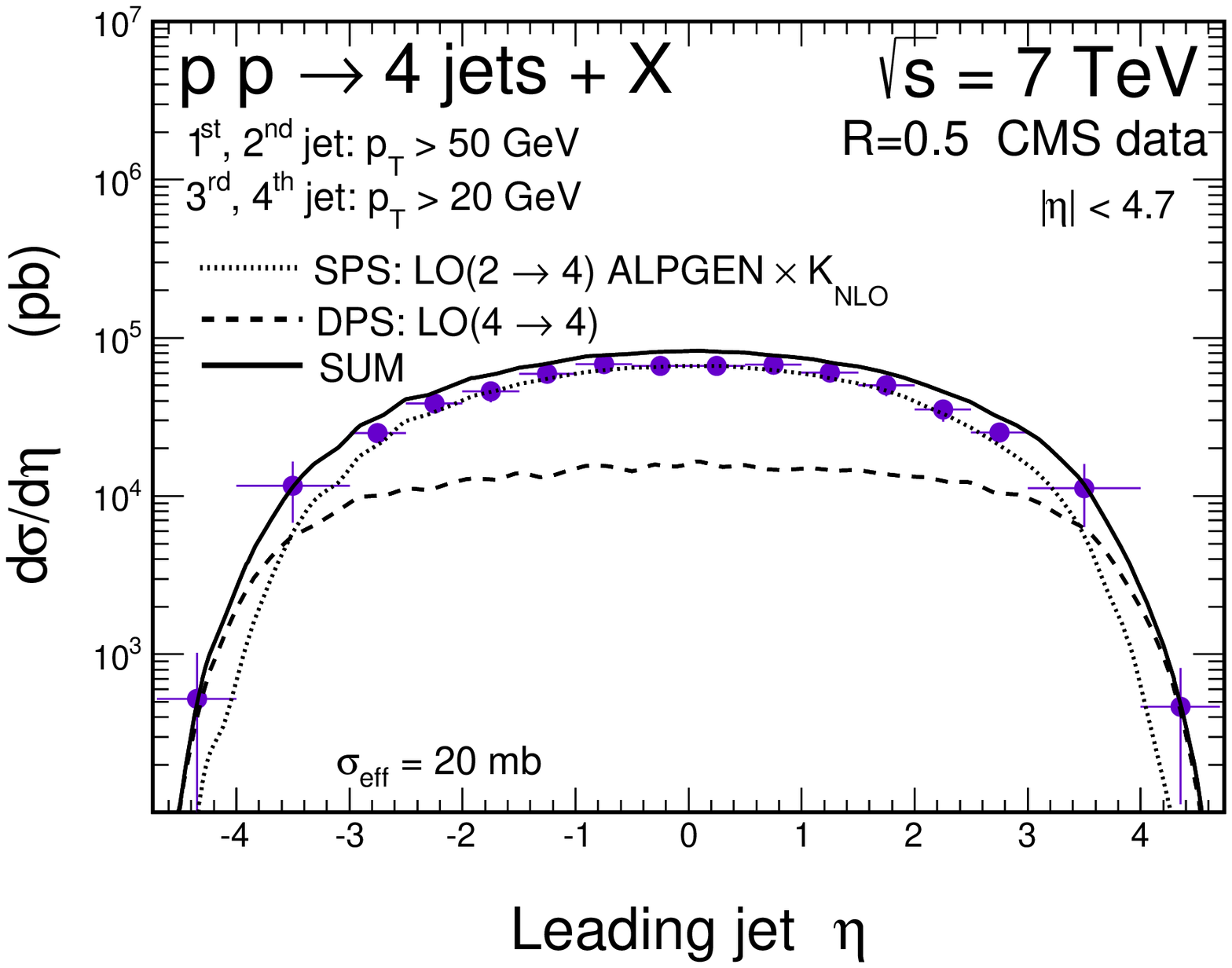}}
\end{minipage}
\begin{minipage}{0.47\textwidth}
 \centerline{\includegraphics[width=1.0\textwidth]{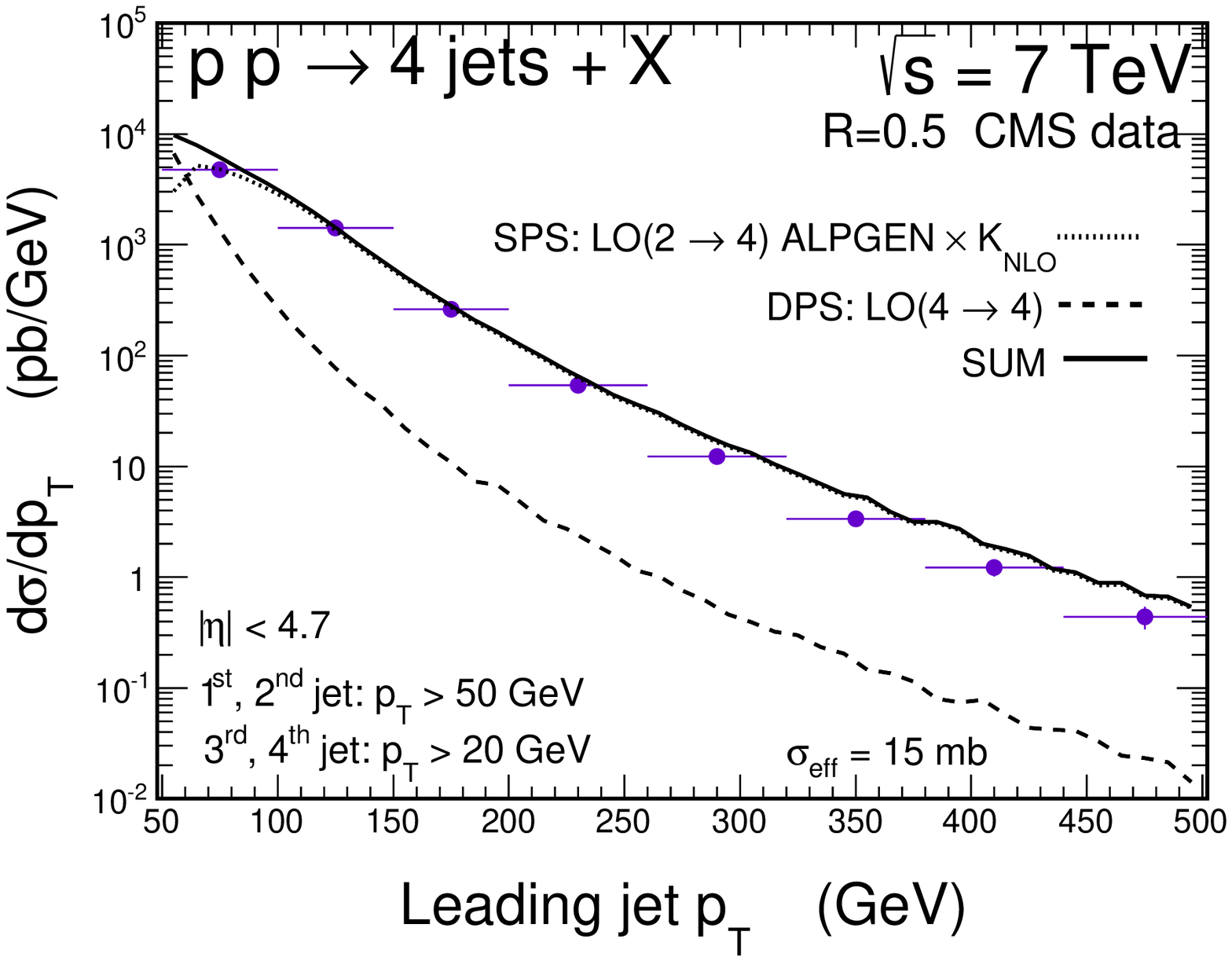}}
\end{minipage}
\hspace{0.5cm}
\begin{minipage}{0.47\textwidth}
 \centerline{\includegraphics[width=1.0\textwidth]{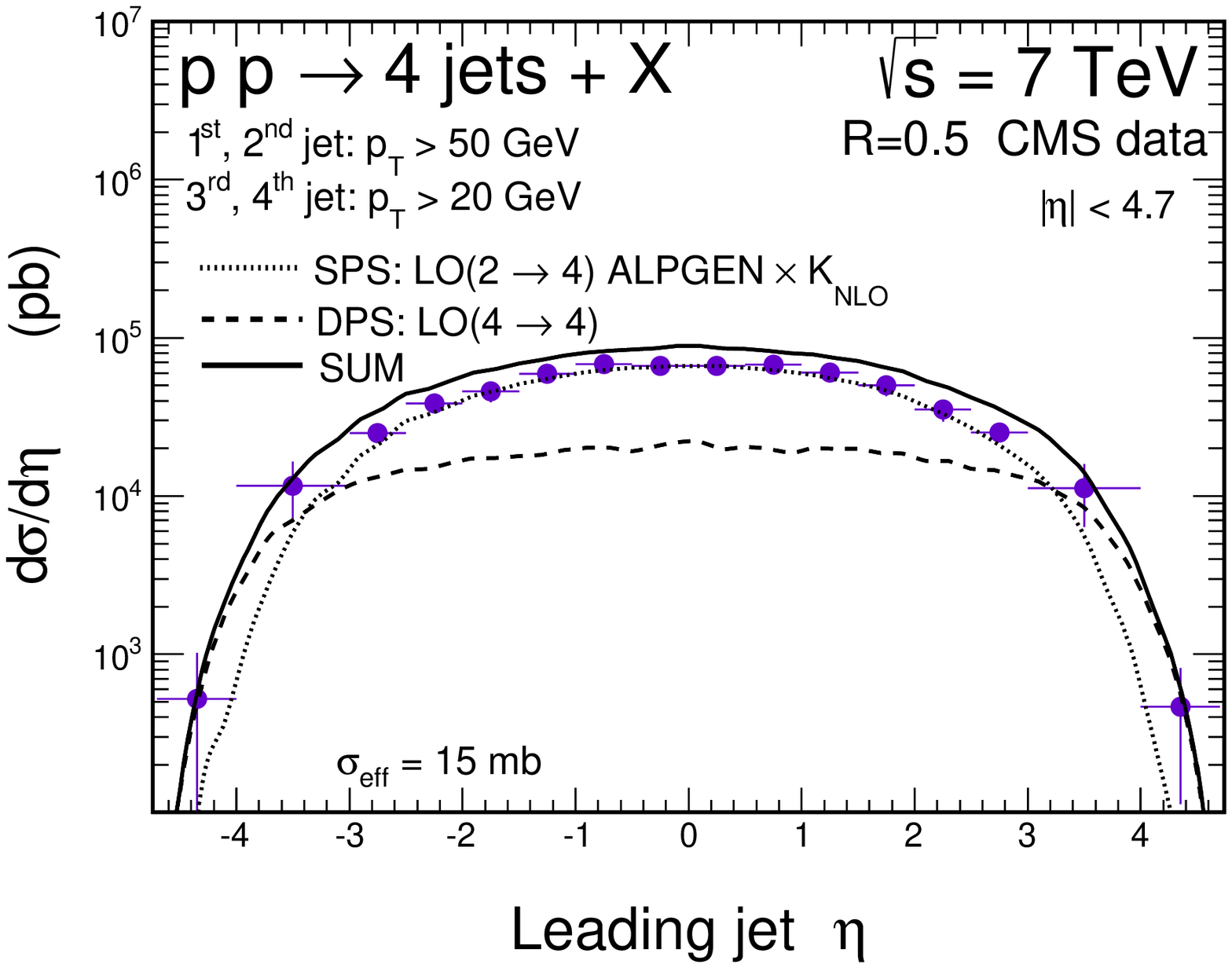}}
\end{minipage}
\begin{minipage}{0.47\textwidth}
 \centerline{\includegraphics[width=1.0\textwidth]{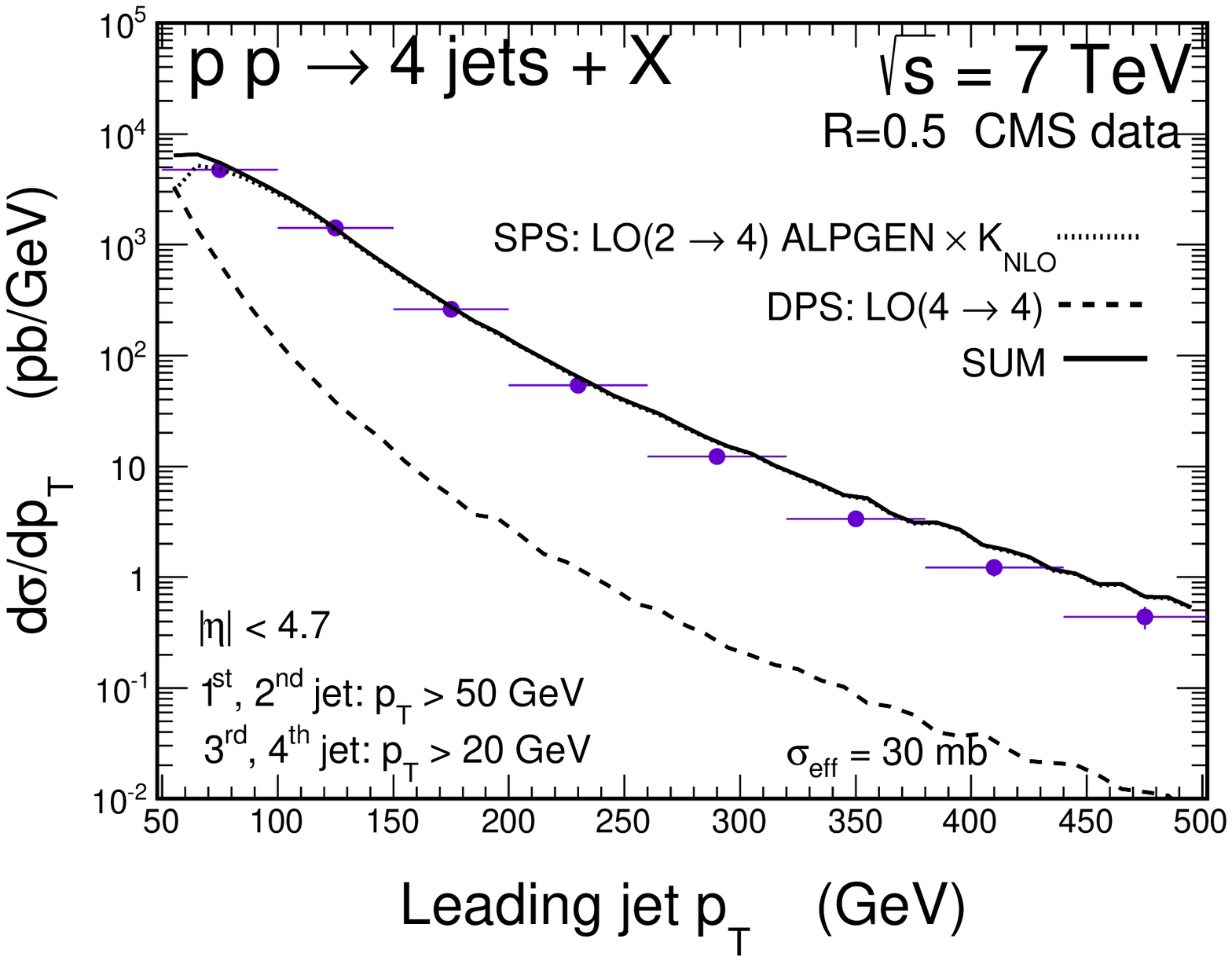}}
\end{minipage}
\hspace{0.5cm}
\begin{minipage}{0.47\textwidth}
 \centerline{\includegraphics[width=1.0\textwidth]{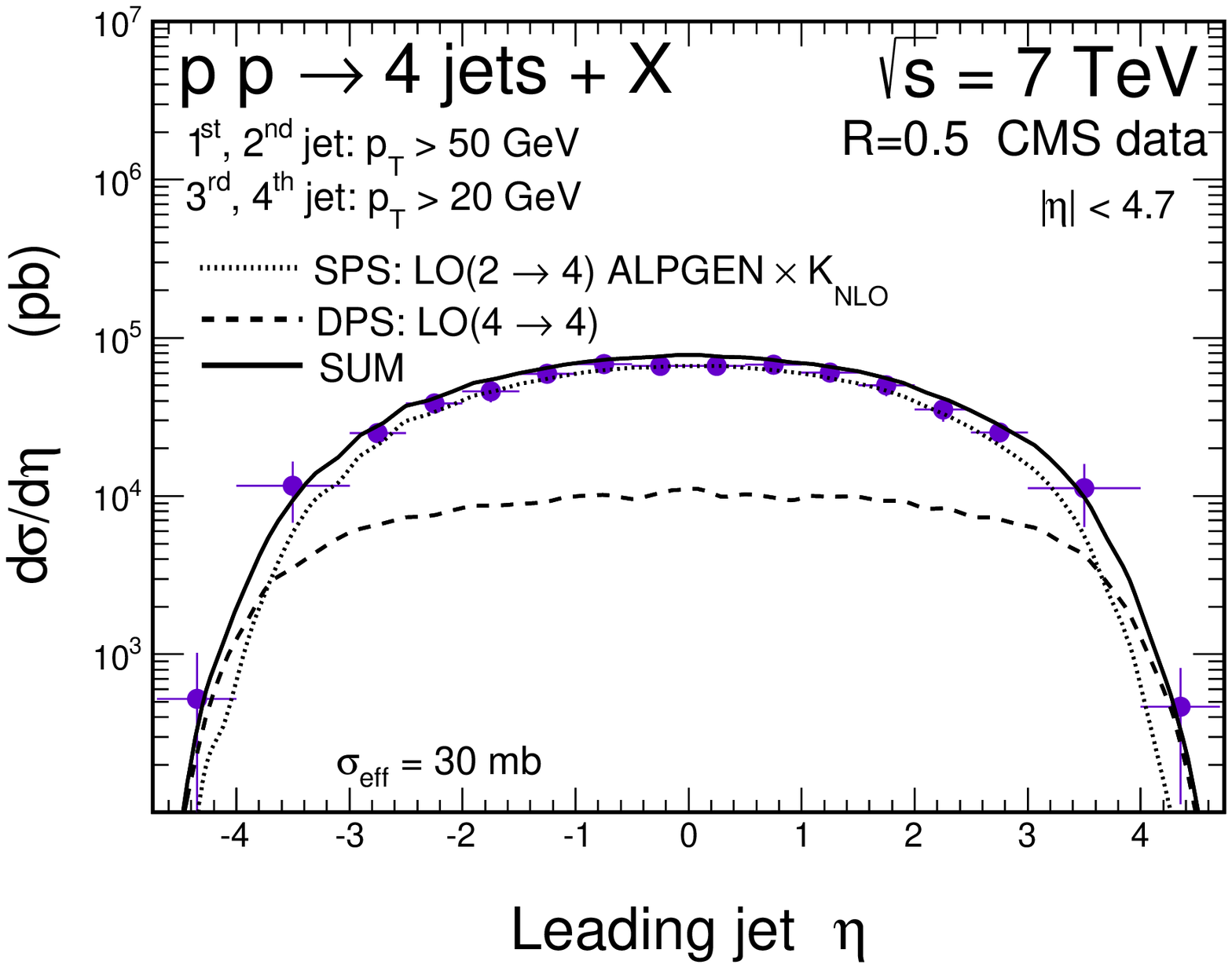}}
\end{minipage}
   \caption{
\small Transverse momentum (left column) and rapidity (right column)
distributions for the leading jet of the four-jet sample.
The solid line represents the SPS four-jet contribution, whereas
the dashed line corresponds to the DPS one.
Different values of $\sigma_{eff}$ (given in the figure) were used in different rows.
}
 \label{fig:CMS-data-2}
\end{figure}

In Fig.~\ref{fig:CMS-data-2} we show in addition the calculated
contributions of DPS for different values of the $\sigma_{eff}$ parameter.
The $\sigma_{eff}$ parameter effectively includes both
traditional (independent emissions) DPS and effects related to
parton splitting \cite{Gaunt2012,Gaunt:2014rua}. In general, the relative contribution
of both mechanisms may depend on different kinematical variables. 
Phenomenological exploration of the dependences may help in
understanding the interplay of the different contributions 
and shed more light on the underlying mechanism.
We observe that the DPS contribution is rather small (independent
of $\sigma_{eff}$ parameter) at larger transverse momenta and in 
midrapidities of jets. Their contribution increases when going to small
transverse momenta and large rapidities.
We observe a clear improvements at large pseudorapidities ($|\eta| >$ 3)
when the DPS contribution is added to the SPS contribution. 
The presented CMS data were not optimized for search for DPS effects
and now we wish to explore how to improve the situation. i.e.
to enhance sample of the DPS events.

\begin{figure}[!h]
\begin{minipage}{0.47\textwidth}
 \centerline{\includegraphics[width=1.0\textwidth]{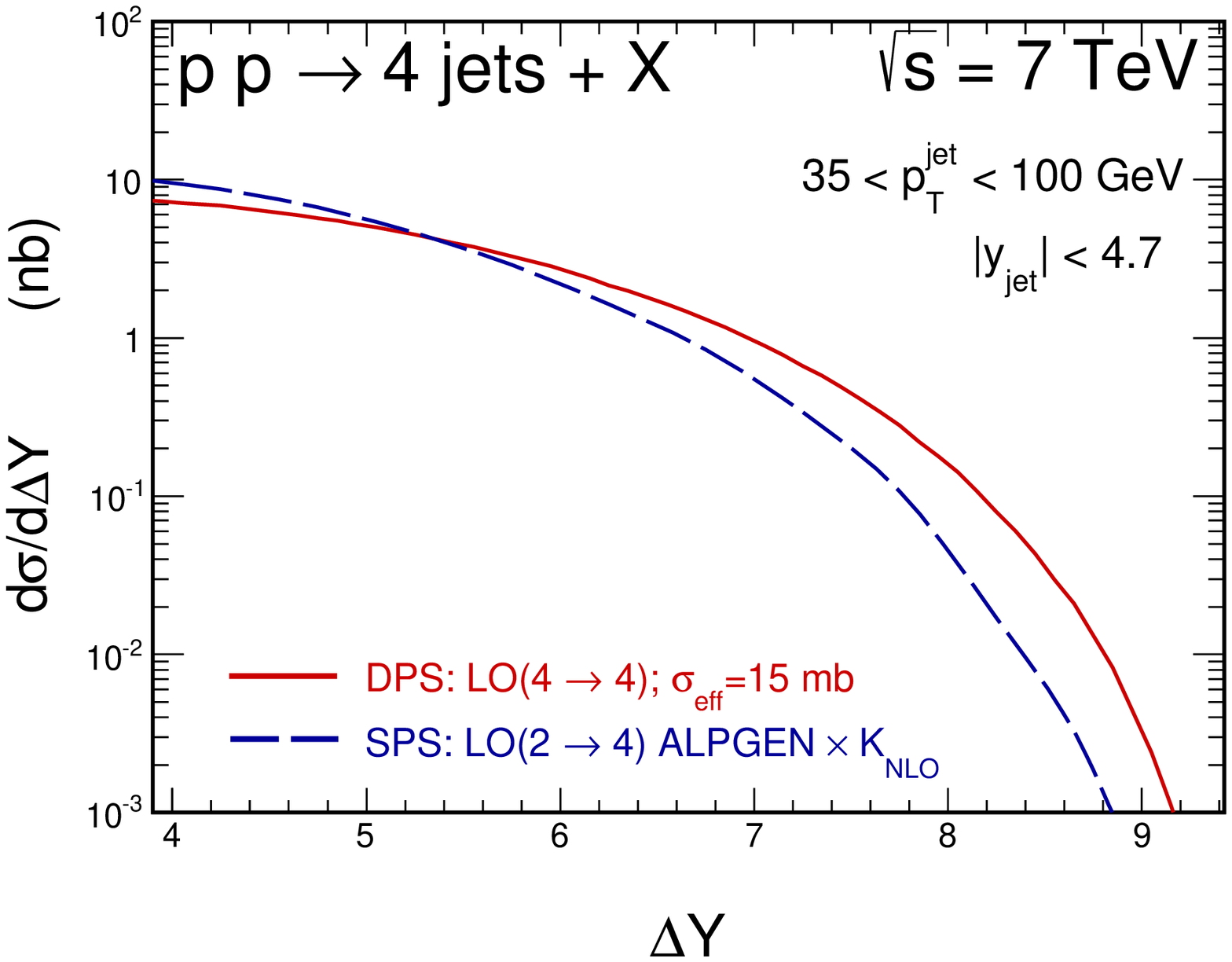}}
\end{minipage}
\hspace{0.5cm}
\begin{minipage}{0.47\textwidth}
 \centerline{\includegraphics[width=1.0\textwidth]{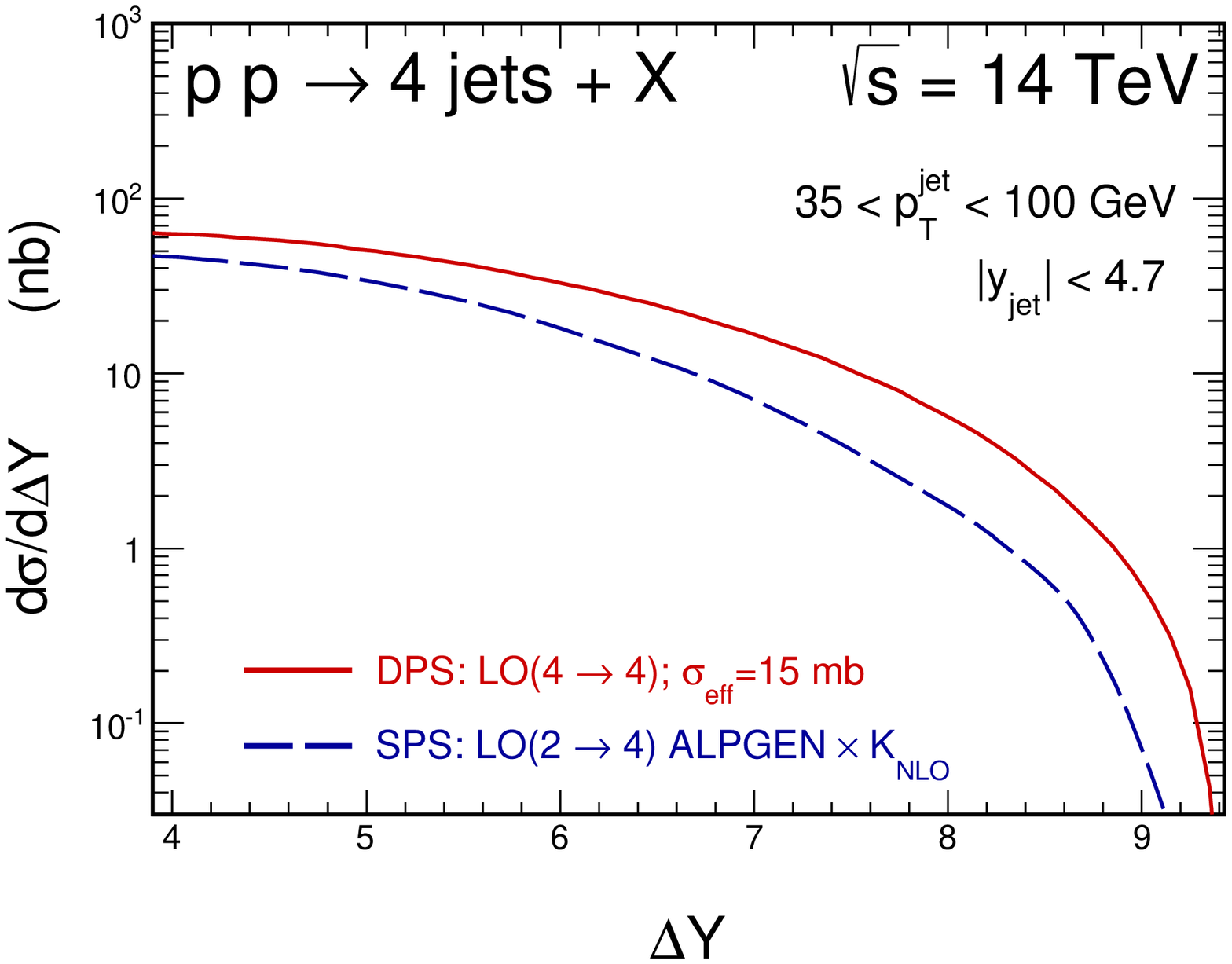}}
\end{minipage}
\begin{minipage}{0.47\textwidth}
 \centerline{\includegraphics[width=1.0\textwidth]{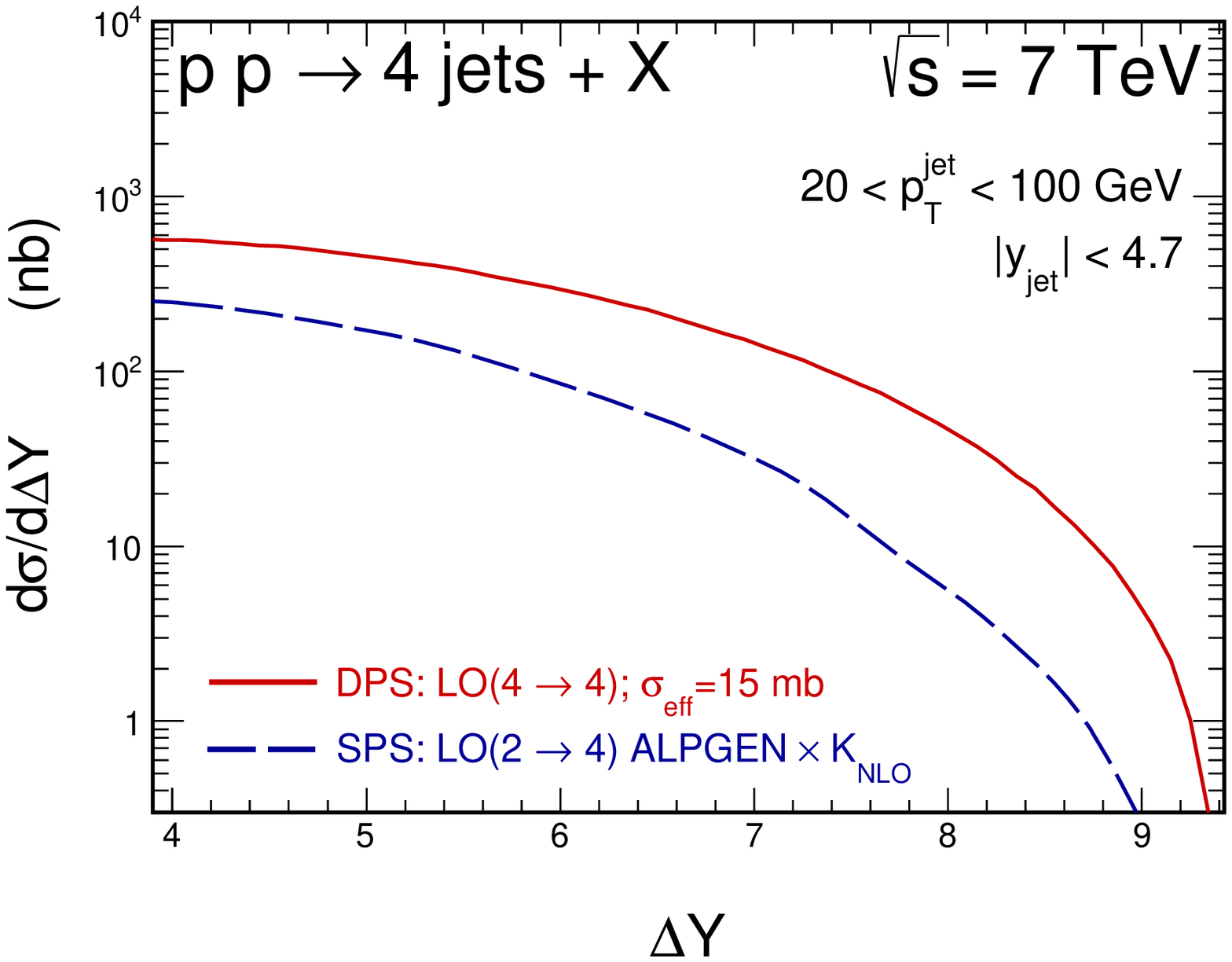}}
\end{minipage}
\hspace{0.5cm}
\begin{minipage}{0.47\textwidth}
 \centerline{\includegraphics[width=1.0\textwidth]{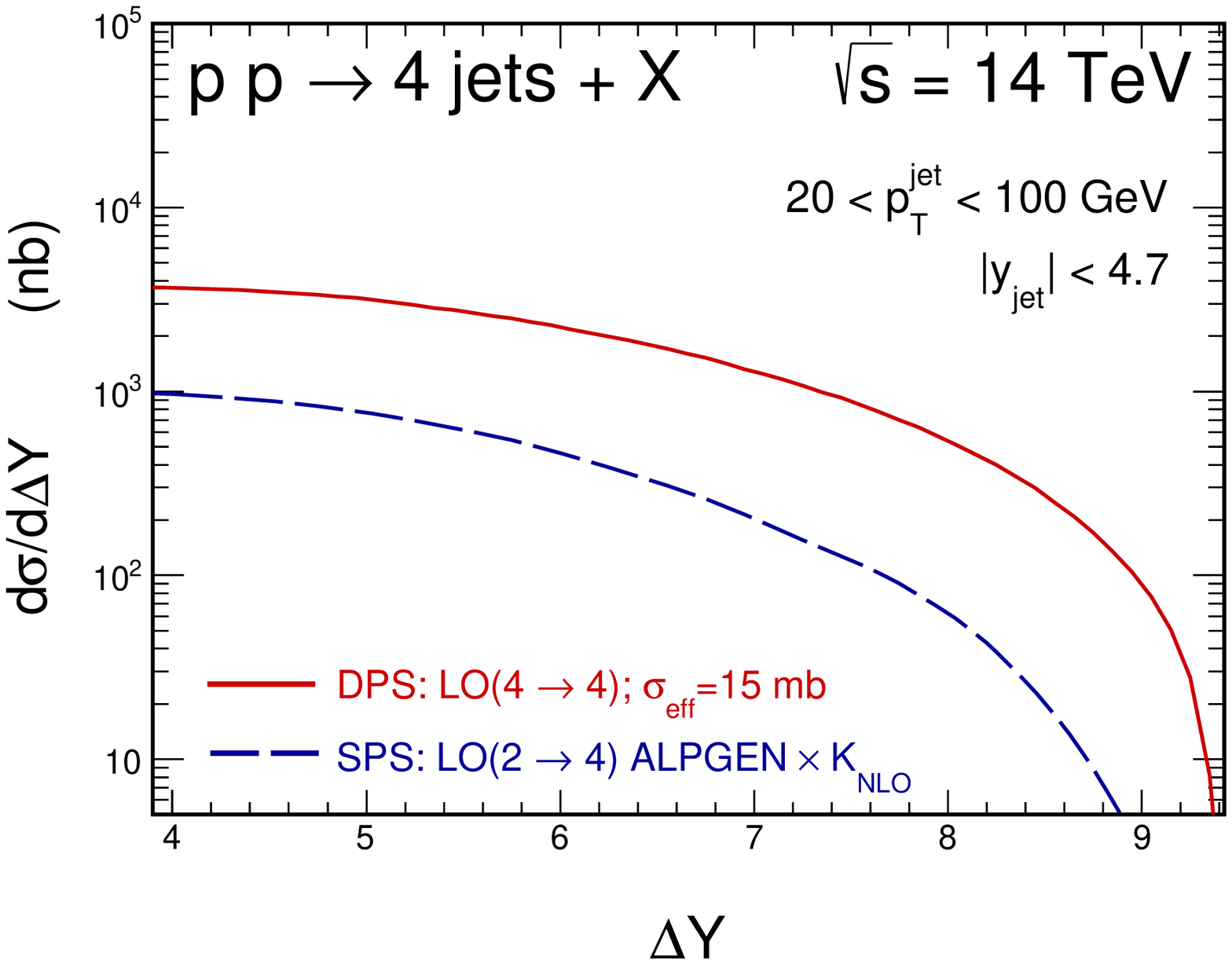}}
\end{minipage}
   \caption{
\small Distribution in rapidity distance of the most remote jets from 
the four-jet sample for $\sqrt{s}$ = 7 TeV (left column) and $\sqrt{s}$
= 14 TeV (right column) for different cuts on jet transverse momenta
(identical for all four jets). }
 \label{fig:DeltaY-DPS-1}
\end{figure}

Having shown that our approach is consistent with existing LHC four-jet data
we can focus on finding optimal conditions for ``observing'' 
the DPS effects. As shown in our previous paper on dijets widely
separated in rapidity \cite{Maciula:2014pla} the distribution in rapidity separation
of such jets seems a very good observable for observing the onset of the DPS enhancement
or even its dominance. In Fig.~\ref{fig:DeltaY-DPS-1} we show some
examples of such distributions for different cuts on the jet transverse
momenta for two collision energies $\sqrt{s}$ = 7 TeV and 
$\sqrt{s}$ = 14 TeV obtained
with the condition of the four-jet observation. We focus only
on the distance between the most remote jets; the other two jets are
then in between. The higher collision energy or the smaller the lower
transverse momentum cut the bigger the relative DPS contribution.
In such cases one can therefore expect a considerable deficit when only
SPS four jets are included. Such cases would be very useful
to "extract" the $\sigma_{eff}$ parameter which for the selected
sample does not to be the same as for other cases discussed in the
literature. Any deviation from the "canonical" value of 15 mb
would therefore shed some new light on the underlying dynamics.
For example, a two-component model discussed in
Refs.~\cite{Gaunt2012,Gaunt:2014rua} strongly suggests such dependences.

\begin{figure}[!h]
\begin{minipage}{0.47\textwidth}
 \centerline{\includegraphics[width=1.0\textwidth]{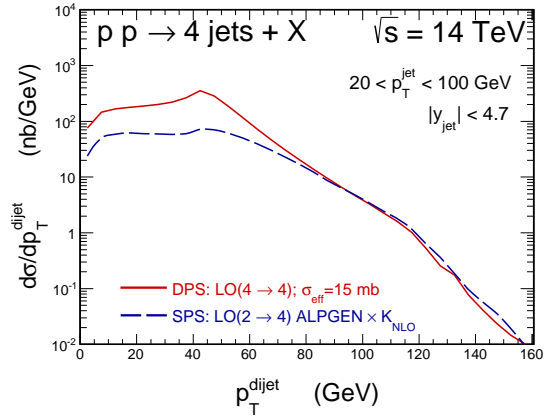}}
\end{minipage}

   \caption{
\small Distribution in transverse momentum of the most remote jets
for $\sqrt{s}$ = 14 TeV.
 }
 \label{fig:ptsum-DPS-1}
\end{figure}

Another observable discussed in Ref.~\cite{Maciula:2014pla} is the distribution
in transverse momentum of the most remote jets.
It was argued there that the jets coming from different 
partonic collisions are not correlated and therefore large transverse
momenta of such uncorrelated jet pairs are possible in general.
Here we wish to explore the situation in this respect quantitatively 
for the four-jet sample. 
In Fig.~\ref{fig:ptsum-DPS-1} we show an example for one 
energy and one cut on jet transverse momenta.
In these calculations we have not included extra cut on jet rapidity
separation. The DPS and SPS distributions look rather similar
for the four-jet sample. This shows that an extra cut on the transverse
momentum of the pair of the most remote jets would almost not help to enhance
the DPS contribution.
 
\begin{figure}[!h]
\includegraphics[width=7cm]{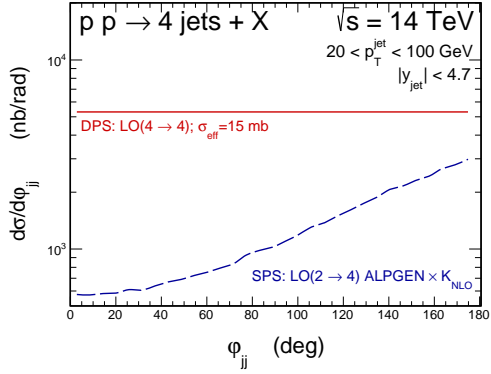}
\caption{
\small Azimuthal correlation between the most remote jets for SPS
(dashed line) and DPS (solid line) contributions for $\sqrt{s}$ = 14 TeV.
}
\label{fig:phi-DPS-1}
\end{figure}

Finally in Fig.\ref{fig:phi-DPS-1} we show an example of azimuthal
angle distribution between the most remote jets.
While the distribution for SPS peaks at $\phi = \pi$, the DPS
contribution is very flat as the most remote jets come dominantly from
different independent and uncorrelated (in azimuthal angle) 
partonic scatterings.
Since the DPS dominates we predict very flat distribution in relative 
azimuthal angle.
Limiting to small $\phi_{jj}$ would further enhance the relative amount
of DPS without considerable lowering statistics.

Table I illustrates and summarizes integrated contributions 
of SPS and DPS for selected transverse momentum cuts imposed on all 
four jets and on rapidity distance between the most remote jets.
Already with the lower cut on transverse momentum
of 35 GeV (used already in some CMS analyses) the DPS contribution
is about 40 \% for $\sqrt{s}$ = 7 TeV and about 60 \% for 
$\sqrt{s}$ = 14 TeV. Lowering the lower cut on jet transverse
momentum to 20 GeV gives already 70\% and 80\% of DPS, respectively. 
Imposing in addition that the distance between the
remote jets is bigger than $\Delta Y >$ 7 enhances the DPS
contribution to almost 90\%. Therefore imposing such cuts would help
to extract fairly precisely the $\sigma_{eff}$ parameter
from such experimental studies. 

\begin{table}[h!]\small%
\caption{Integrated cross sections in nanobarns
for two collision energies and different cuts on jet transverse momenta
and rapidity distance between the most remote jets. Here, $\sigma_{eff} = 15$ mb has been used for calculating the DPS cross section.}
\newcolumntype{Z}{>{\centering\arraybackslash}X}
\label{table}
\centering %
\begin{tabularx}{16.cm}{c|ZZc|ZZc}
\toprule[0.1em] %

Kinematical cuts:   & \multicolumn{3}{c|}{$\sqrt{s} = 7$ TeV} & \multicolumn{3}{c}{$\sqrt{s} = 14$ TeV} \\  
$|y| < 4.7$        & $\sigma^{SPS}$  & $\sigma^{DPS}$ & {\scriptsize $\frac{DPS}{SPS+DPS}$} &$\sigma^{SPS}$ & $\sigma^{DPS}$ & {\scriptsize $\frac{DPS}{SPS+DPS}$}    \\   
\toprule[0.1em]

\multirow{2}{4.5cm}{$35 < p_{T} < 100$ GeV} & \multirow{2}*{40.55}  & \multirow{2}*{29.92} & \multirow{2}*{42\%} & \multirow{2}*{197.74} & \multirow{2}*{275.23}  & \multirow{2}*{58\%}  \\
& & & & & &\\
\hline
\multirow{2}{4.5cm}{$20 < p_{T} < 100$ GeV} & \multirow{2}*{1 047.37}  & \multirow{2}*{2 443.77} & \multirow{2}*{70\%} & \multirow{2}*{4 194.11} & \multirow{2}*{16 652.39} & \multirow{2}*{80\%} \\ 
& & & & & & \\
\hline
\multirow{1}{4.5cm}{$20 < p_{T} < 100$ GeV} & \multirow{2}*{18.56}  & \multirow{2}*{113.26} & \multirow{2}*{86\%} & \multirow{2}*{151.70} & \multirow{2}*{1 194.28}   & \multirow{2}*{89\%}  \\                
\multirow{1}{4.5cm}{$\Delta Y > 7.0\;$}                           &  &  &  & & &\\
\hline
\multirow{1}{4.5cm}{$20 < p_{T} < 100$ GeV} & \multirow{2}*{291.68}  & \multirow{2}*{1 221.88} & \multirow{2}*{81\%} & \multirow{2}*{1 157.15} & \multirow{2}*{8 326.19} & \multirow{2}*{88\%}   \\                
\multirow{1}{4.5cm}{$0< \varphi_{jj} < \frac{\pi}{2}\;$}                           & & &  &  &  & \\

\bottomrule[0.1em]

\end{tabularx}
\end{table}

\section{Conclusions}

In the present paper we have explored how to enhance the relative
contribution of double-parton scattering for four-jet production.
Compared to our previous studies, where we focussed on jets with
large rapidity separation (important in searches for the BFKL dynamics),
here we have studied the case of exclusive four-jet production. 
We have shown that already some present data for four-jet production 
obtained at the LHC by the CMS collaboration
with relatively small cuts on transverse momenta indicate some evidence 
of DPS at large pseudorapidities of the leading jet.

We have shown that imposing a lower cut on transverse momenta and 
rapidity distance between the most remote jets improves 
the situation considerably, i.e.enhances the relative contribution of DPS.
A dedicated analysis of the DPS effect is possible already with 
the existing data sample at $\sqrt{s}$ = 7 TeV. The situation 
at larger energies, relevant for LHC Run 2 should be even better.
We predict that azimuthal correlation between jets widely separated
in rapidity should dissapear in the considered kinematical domain.
We have found that in some corners of the phase space the DPS
contribution can go even above 80 \%, not necesserily at the 
expense of lowering the cross section (statistics of experimental data).

We have presented the detailed predictions. Once such cross sections 
are measured, one could try to extract the $\sigma_{eff}$ parameter
and try to obtain its dependence on kinematical variables.
Such dependence can be expected due to several reasons 
(parton-parton correlations, hot spots, perturbative parton splitting)
and detailed experimental data would help us to understand
the situation much better.

\vspace{1cm}

{\bf Acknowledgments}
The work of A.S. was partially supported by the Centre for Innovation
and Transfer of Nautral Sciences and Engineering Knowledge
in Rzesz\'ow.



\end{document}